\documentclass[summary]{ursi}
\usepackage[super]{nth}

\title{Evidence of changes in the low-latitude plasma drift under IMF $B_z$ coupling: a TIEGCM simulation approach}

\author{Sumanjit Chakraborty}

\affiliation{
  Space and Atmospheric Sciences Division, Physical Research Laboratory, Ahmedabad 380009, Gujarat, India
}

\begin{document}

\maketitle

\begin{abstract}

Study of the dynamic nature of low-latitude ionosphere during geomagnetically disturbed conditions, especially in the EIA and the magnetic equatorial regions are vital for understanding the underlying physics as well as for mitigating space weather hazards on the sophisticated technological systems essential for human civilization. An important aspect of the space weather studies is the thorough understanding of coupling between the solar wind and the terrestrial magnetosphere-ionosphere system and subsequent influence on the low-latitude ionosphere. This paper presents an effort to understand the influence of actual values of the north-south component of Interplanetary Magnetic Field (IMF, $B_z$) on the vertical plasma drifts at a location near the EIA and the geomagnetic equator. The strong storm event of October 13, 2016, falling in the descending phase of solar cycle 24, has been taken up as a case study. Thermosphere-Ionosphere-Electrodynamics General Circulation Model (TIEGCM) simulation runs have been performed under two scenarios: first when no coupling is present (IMF $B_z$ = 0 nT) and second when actual observations of the values of IMF $B_z$ is given as inputs to the model. Observations show when actual data is fed to the model, there is significant shift of the vertical drift towards westward, while there is an increase in the peak value of the westward drift after the pre-reversal enhancement. This study is an initial effort to understand the variations in low-latitude plasma motions during the main phase of strong geomagnetic storms. This initial work will be followed up with understanding the global plasma drift variations under the influence of other components of the IMF near the EIA and the dip equatorial regions.          
\end{abstract}

\section{Introduction}

The occurrence of geomagnetic storms is associated with input of the energy of the solar wind and subsequent coupling of the same with the Magnetosphere-Ionosphere (MI) system. In general, these storms occur when the north-south component ($B_z$) of the frozen-in Interplanetary Magnetic Field (IMF), emanating from the Sun and carried by the solar wind into the heliosphere, turns completely southward (or negative $B_z$ by convention) and remains in this state for several hours while reconnecting with the geomagnetic field [1-2]. These phenomena occur mainly following a Coronal Mass Ejection (CME) or successive CMEs and the type of magnetic storms driven by this solar transient are generally refereed to as CME-driven geomagnetic storms. Additionally, these type of storms occur from components like the strong magnetic field in the ejecta, the strong magnetic field of the sheath and the interplanetary shocks. It is to be noted that these storms mainly occur in the ascending phase and solar maximum phase of solar cycles (or solar activity cycle of $\approx$ 11 years duration). The Storm Sudden Commencements (SSC), before the onset of the main phase of these type of storms, are associated with the interplanetary shocks that are driven by the CMEs while the recovery phase lasts for only about one-two days [3-4]. A geomagnetic storm's beginning is noted with large variations of $B_z$ in addition to the transmission of magnetospheric convection electric field (with an associated equivalent current system having average periods of the order of few minutes to about 3 hours) from the higher to the lower-equatorial latitudes, also called as the Prompt Penetration Electric Fields (PPEF and eastward during dawn-to-dusk) which cause severe perturbations to the low-latitude electrodynamics [5-10].

The ionosphere of the low-latitude region is characterized by a phenomenon known as the Equatorial Ionization Anomaly (EIA), that starts around 09 Local Time (LT) and moves towards higher latitudes up to 15-20${^\circ}$ magnetic dip latitude around 16 LT and returns back to the magnetic equator around 24 LT. This phenomena is caused as a result of the two plasma motions: one perpendicular to the geomagnetic field and generating drifts upward the zonal electric field (eastward during daytime and westward during nighttime) produced by E region dynamo and the geomagnetic field and the other parallel to the geomagnetic field, that cause the plasma to diffuse down following the geomagnetic field line under influence of ambipolar diffusion drift related to gravity and pressure gradients [11-16].         

The present work investigates the changes in the plasma drift, near the dip equator and the EIA regions, that were caused due to the influence of the CME-driven geomagnetic storm of October 13, 2016 by utilizing simulations from the Thermosphere-Ionosphere-Electrodynamics General Circulation Model (TIE-GCM) developed by NCAR. Firstly, the analysis is performed by simulating the vertical plasma drifts at Indore:IDR (a location near the EIA, magnetic dip of  32.23${^\circ}$N) and Tirunelveli: TIR (a dip equatorial location, magnetic dip of 0.2${^\circ}$N) under no variations of the IMF $B_z$ (i.e: $B_z$ = 0 nT) and then by simulating the same for these two locations, with observed values of $B_z$ during the entire day of October 13, 2016. The motivation of the work comes from the need of understanding these physics-based models and their performance such that these models can be reliably run across locations where actual data from ionosondes or GNSS receivers are not present. The paper is divided starting with a general discussion of the NCAR TIEGCM model followed by the results, the corresponding discussions and the summary sections.

\section{Data: The TIEGCM}

The National Center for Atmospheric Research (NCAR) TIE-GCM (https://www.hao.ucar.edu/modeling/tgcm/tie.php) is a 3D, non-linear and first principles representation of the coupled ionosphere and thermosphere system. It consists of self-consistent solution of the mid-and low-latitude dynamo field. It solves energy, continuity and momentum equations for ion and neutral species at every time-step (typically 120 s) using a semi-implicit, $\nth{4}$-order, finite difference method on each pressure surface in a vertical grid. The standard parameters in this model are: latitudinal and longitudinal extent from -87.5${^\circ}$ to +87.5${^\circ}$ and -180${^\circ}$ to +180${^\circ}$ respectively, in steps of 5${^\circ}$, altitude-wise pressure level from -7 to +7 in steps of $\frac{H}{2}$ with lower boundary $\approx$97 km and upper boundary $\approx$500-700 km (depending on the solar activity condition). The inputs to this model are: daily solar radio flux (F10.7) and its 81-day centered mean (F10.7$\_$81) while the output parameters (specified in 3D spatial and time dimensions) are: Height of pressure surfaces in cm, electron, ion and neutral temperatures in K, compositions like N2+, NO+, N+, Ne,  NO, N(4S), N(2D), O, O2, O+, O2+, meridional, vertical and zonal drifts in m/s and potentials both geographic and geomagnetic coordinates. Several researchers [17-22] have shown the capability of the TIEGCM to understand geomagnetically disturbed as well as quiet-time conditions' ionospheric electrodynamic processes thus making it an ideal simulation platform that closely reproduces and resembles the same. 

\section{Results and Discussions}

\subsection{Interplanetary and geomagnetic conditions}

\begin{figure}[htbp]
\centering
\includegraphics[width=85mm]{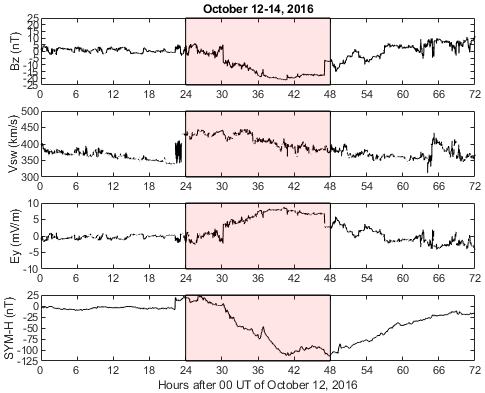}
\caption{From top to bottom: IMF $B_z$ (in nT), Vsw (in km/s), IEF $E_y$ (mV/m) and the SYM-H (in nT) from October 12 through 14, 2016. The shaded rectangle (24-48 UT) marks October 13, 2016: the period of main phase of the geomagnetic storm.}
\label{sc01}
\end{figure}

A CME erupted into the space as a result of the eruption of a magnetic filament from the northern hemisphere of the Sun around 16 UT on October 08, 2016. The CME arrived at L1-point around 21:21 UT on October 12, 2016. Following the events, a G2 class ($K\_p$ = 6, NOAA scales: http://swpc/noaa.gov/noaa) geomagnetic storm commenced on October 13, 2016 at around 08:15 UT (https://www.spaceweather.com/archive). Figure \ref{sc01} from top panel to bottom panel shows the IMF $B_z$ (in nT), the average solar wind flow speed Vsw (in km/s), the Interplanetary Electric Field (IEF) $E_y$ (in mV/m) and the one-minute average of Dst index: SYM-H index (in nT) during October 12-14, 2016. The shaded region (24-48 UT) shows the day (October 13, 2016) of the main phase of the geomagnetic storm. The IMF $B_z$ had been below -10 nT during October 13, 2016 from 08:26 UT to 21:58 UT. It reached minimum value of -21 nT on 15:18 UT of the same day, thus signifying strong coupling between the solar wind and the MI system. The average Vsw values on October 12, 2016 had been around 367 km/s, rose to values around 411 km/s during the day of the storm main phase (October 13, 2016) and dropped back to about 369 km/s on October 14, 2016. The IEF $E_y$ had also showed a peak on October 13, 2016 around 16:18 UT with a value of 8.77 mV/m, as a result of higher values of Vsw and strong southward IMF $B_z$ conditions on October 13, 2016. The SYM-H here shows minimum value of -114 nT on 23:45 UT, thus designating the event to be a strong/intense (Dst/SYM-H $<$-100 nT) geomagnetic storm [23]. 

\subsection{TIEGCM results}

\begin{figure}[htbp]
\centering
\includegraphics[width=85mm]{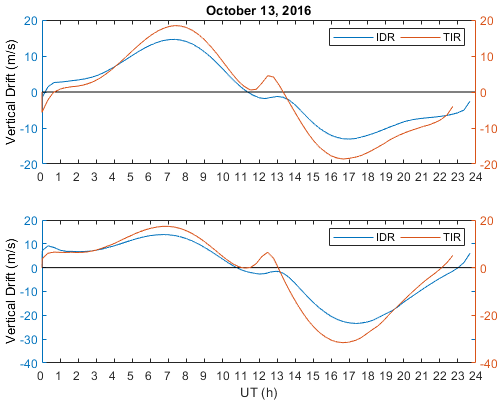}
\caption{Variation of the vertical plasma drifts over an equatorial location (TIR in red) and an off-equatorial location (IDR in blue) on the day (October 13, 2016) of main phase of the geomagnetic storm event, when Top panel (Scenario 1): IMF $B_z$ is set to zero throughout the TIEGCM run and bottom panel (Scenario 2): when actual values of $B_z$ is fed as inputs to the model throughout the main phase period.}
\label{sc02}
\end{figure}

Figure \ref{sc02} shows the TIEGCM simulation results on October 13, 2016. As discussed in the introduction section, the locations observed are IDR and TIR which are locations near the EIA and the geomagnetic equator passing the Indian subcontinent. The top panel shows the vertical drift variation of both IDR (in blue) and TIR (in red) when IMF $B_z$ was set to 0 nT (scenario 1). The bottom panel shows the same with the observed IMF $B_z$ values as obtained from the omniweb database (scenario 2). On investigating the simulation results, it can be seen that when the model is fed with actual or observed data, the difference in the eastward (positive values) peak of the vertical plasma drift reduces while there is an increase of the same in the westward (negative values) peak. Furthermore, feeding of the observed IMF $B_z$ reduces the overall level of the eastward drift but increases the westward drift after the Pre-Reversal Enhancement (PRE) around 13 UT (or 18:30 LT). It is also to be noted that at an off-equatorial location (IDR), the PRE gets suppressed in both the simulation scenarios, however there is no effect of the same on equatorial location's (TIR) PRE levels. The possible reason of these suppressions could be attributed to the disturbance dynamo effect which is getting reflected in an off-equatorial location. From this figure it is evident that when applying real data of the IMF $B_z$, the overall level of the drift is getting reduced (more westward). Thus indicating a possible mechanism that is affecting the vertical drifts of both off-equatorial and equatorial locations under geomagnetic storm effect.           

\section{Summary}

The low-latitude ionosphere is dynamic in nature, especially in and around the EIA, where almost two-thirds of the global ionization gets concentrated. In addition, the perturbations might increase or decrease when there is an external driver such as geomagnetic storms as a result of solar outburst and subsequent strong coupling of the solar wind and MI system. There arises a need to understand the effect of the IMF parameters (especially $B_z$) during such events on the plasma motion in the form of drift from equatorial to off-equatorial locations. This study compares the TIEGCM simulation results of the vertical plasma drift over both EIA and magnetic equator in the Indian sector under the scenario of no coupling of the solar wind with MI system ($B_z$ = 0 nT) and a scenario with actual observations of $B_z$ fed to this model. Observations brought forward the fact that there had been significant difference in the variation of the vertical plasma drift, in the form of suppression of the entire profile towards westward, when the model was fed with actual observations of IMF $B_z$. Further studies with variations of the IMF conditions, under geomagnetically disturbed as well as quiet-time conditions, fed to the TIEGCM model could bring out new physical mechanisms that are taking place at such highly geosensitive locations over low-to-equatorial latitudes, especially in and around the EIA and the magnetic equator. This initial work will be followed up by thorough investigation of the effects of the other IMF components on the global plasma drift motions over these dynamic locations.

\section{Acknowledgements}

SC acknowledges the World Data Center (WDC) for Geomagnetism, Kyoto for the Dst index value. The solar wind and interplanetary magnetic field and the SYM-H index data are obtained from NASA’s SPDF omniweb service https://omniweb.gsfc.nasa.gov/form/omni\_min.html. This work is supported by the Department of Space, Government of India.


\begin{thebibliography}{99}

\bibitem{1} J.W. Dungey. Interplanetary magnetic field and the auroral zones. Phys. Rev. Lett., 6 (1961), pp. 47-48, 10.1103/PhysRevLett. 6.47

\bibitem{2} W.D. Gonzalez, J.A. Joselyn, Y. Kamide, H.W. Kroehl, G. Rosoker, B.T. Tsurutani, V.M. Vasyliunas. What is a geomagnetic storm? J. Geophys. Res., 99 (A4) (1994), pp. 5771-5792, 10.1029/93JA02867

\bibitem{3} J.E. Borovsky, M.H. Denton. Differences between CME-driven storms and CIR-driven storms. J. Geophys. Res., 111 (2006), p. A07S08, 10.1029/2005JA011447

\bibitem{4} S. Chakraborty, S. Ray, D. Sur, A. Datta, A. Paul. Effects of cme and cir induced geomagnetic storms on low-latitude ionization over indian longitudes in terms of neutral dynamics. Adv. Space Res., 65 (2020), pp. 198-213, https://doi.org/10.1016/j.asr.2019.09.047

\bibitem{5} A. Nishida. Geomagnetic 2 fluctuations and associated magnetospheric phenomena. J. Geophys. Res., 73 (5) (1968), pp. 1795-1803, 10.1029/JA073i005p01795

\bibitem{6} V.M. Vasyliunas. Mathematical models of the magnetospheric convection and its coupling to the ionosphere. McCormacMcCormac (Ed.), Particles and Fields in the Magnetosphere, Springer, New-York (1970), pp. 60-71, 10.1007/978-94-010-3284-1-6

\bibitem{7} B.G. Fejer, M.F. Larsen, D.T. Farley. Equatorial disturbance dynamo electric fields. Geophys. Res. Lett., 10 (1983), pp. 537-540, 10.1029/GL010i007p00537

\bibitem{8} R.W. Spiro, R.A. Wolf, B.G. Fejer. Penetration of high latitude electric field effects to low latitudes during the SUNDIAL 1984. Ann. Geophys., 6 (1988), pp. 39-49

\bibitem{9} B.G. Fejer, L. Scherliess. Mid-and low-latitude prompt penetration ionospheric zonal plasma drifts. Geophys. Res. Lett., 25 (16) (1997), pp. 3071-3074, 10.1029/98GL02325

\bibitem{10} S. Basu, Su. Basu, E. MacKenzie, C. Bridgwood, C.E. Valladares, K.M. Groves, C. Carrano. Specification of the occurrence of equatorial ionospheric scintillations during the main phase of large magnetic storms within solar cycle. Radio Sci., 45 (5) (2010), p. RS5009, 10.1029/2009RS004343

\bibitem{11} R.G. Rastogi. The diurnal development of the anomalous equatorial belt in the F2 region of the ionosphere. J. Geophys. Res., 64 (1959), pp. 727-732, 10.1029/JZ064i007p00727

\bibitem{12} P.C. Kendall, W.M. Pickering. Magnetoplasma diffusion at F2-region altitudes. Planet. Space Sci., 15 (1967), pp. 825-833, 10.1016/0032- 0633(67)90118-3

\bibitem{13} D.N. Anderson. A theoretical study of the ionospheric F-region equatorial anomaly, II, results in the American and Asian sectors. Planet. Space Sci., 21 (42) (1973), pp. 421-442

\bibitem{14} H. Rishbeth. The equatorial. F-layer: progress and puzzles Ann. Geophys., 18 (2000), pp. 730-739, 10.1007/s00585-000-0730-6

\bibitem{15} Chakraborty, S., Datta, A., Ray, S., Ayyagari, D., and Paul, A. (2020b). Comparative studies of ionospheric models with gnss and navic over the indian longitudinal sector during geomagnetic activities. Advances in Space Research,66(4), 895-910, https://doi.org/10.1016/j.asr.2020.04.047 

\bibitem{16} Chakraborty, S., Ray, S., Datta, A., and Paul, A.(2020c). Ionospheric response to strong geomagnetic storms during 2000–2005: An imf clock angle perspective. Radio Science,55,e2020RS00706, https://doi.org/10.1029/2020RS007061 

\bibitem{17} Richmond, A. D., Peymirat, C., and Roble, R. G. (2003), Long-lasting disturbances in the equatorial ionospheric electric field simulated with a coupled magnetosphere-ionosphere-thermosphere model, J. Geophys. Res., 108, 1118, 10.1029/2002JA009758, A3.

\bibitem{18} Vichare, G., and Richmond, A. D. (2005), Simulation study of the longitudinal variation of evening vertical ionospheric drifts at the magnetic equator during equinox, J. Geophys. Res., 110, A05304, 10.1029/2004JA010720

\bibitem{19} Rodrigues, F. S., Crowley, G., Heelis, R. A., Maute, A., and Reynolds, A. (2012), On TIE-GCM simulation of the evening equatorial plasma vortex, J. Geophys. Res., 117, A05307, 10.1029/2011JA017369

\bibitem{20} Richmond, A. D., Fang, T. -W. ., and Maute, A. (2015), Electrodynamics of the equatorial evening ionosphere: 1. Importance of winds in different regions. J. Geophys. Res. Space Physics, 120, 2118– 2132. 10.1002/2014JA020934

\bibitem{21} Maute, A., Richmond, A. D., Lu, G., Knipp, D. J., Shi, Y., $\&$ Anderson, B. (2021). Magnetosphere-ionosphere coupling via prescribed field-aligned current simulated by the TIEGCM. Journal of Geophysical Research: Space Physics, 126, e2020JA028665. 10.1029/2020JA028665

\bibitem{22} Shiokawa, K., Lu, G., Otsuka, Y., Ogawa, T., Yamamoto, M., Nishitani, N., and Sato, N. (2007), Ground observation and AMIE-TIEGCM modeling of a storm-time traveling ionospheric disturbance, J. Geophys. Res., 112, A05308, 10.1029/2006JA011772

\bibitem{23} C.A. Loewe, G.W. Prolss. Classification and mean behavior of magnetic storms. J. Geophys. Res.: Space Phys., 102 (1997), pp. 14209-14213, 10.1029/96JA04020


\end{thebibliography}
\end{document}